\documentclass[a4paper,twocolumn,11pt,unpublished]{quantumarticle}

\pdfoutput=1\usepackage[utf8]{inputenc}   % pdfLaTeX only

\usepackage{float}
\usepackage[ruled,vlined,linesnumbered]{algorithm2e}
\usepackage{pgfplots}
\pgfplotsset{compat=1.18}

\usepackage[english]{babel}
\usepackage[T1]{fontenc}
\usepackage{amsmath}
\usepackage{tikz}
\usepackage{lipsum}

\usepackage{graphicx}
\usepackage{amsfonts, amssymb, amsthm}
\usepackage{xcolor}
\usepackage{braket}

\usepackage[compat=0.3]{yquant}

\usepackage{thm-restate}

\newtheorem{theorem}{Theorem}
\newtheorem{corollary}{Corollary}
\newtheorem{lemma}{Lemma}

\newtheorem{definition}{Definition}
\newcommand{\opnorm}[1]{\left|\left|#1\right|\right|_{op}}
\newcommand{\linfnorm}[1]{\left|\left|#1\right|\right|_{\infty}}

\newcommand{\cost}[1]{\mathcal{O}\left(#1\right)}

\usepackage{rotating}

\usetikzlibrary{arrows.meta, positioning}
\usepackage{adjustbox}
\usepackage[numbers]{natbib}

\definecolor{c1}{RGB}{44, 62, 80}
\definecolor{darkgray176}{RGB}{176,176,176}
\definecolor{c3}{RGB}{58, 125, 68}
\definecolor{c2}{RGB}{125, 128, 127}
\definecolor{c4}{RGB}{164, 166, 46}
\definecolor{c5}{RGB}{212, 180, 76}
\definecolor{c6}{RGB}{44, 62, 80}
\definecolor{c7}{RGB}{58, 125, 68}

\usepackage{hyperref}
\usepackage{cleveref} 
\begin{document}

\title{von Neumann measurement and quantum phase estimation of block-encoded Hamiltonians}
 \author{S. E. Skelton}
 \email{shawn.skelton@itp.uni-hannover.de}
 \affil{Institute for Theoretical Physics, University of Leibniz, Hannover Hannover, Germany}
\maketitle
\begin{abstract}
We review how to use von Neumann's measurement procedure to estimate a phase, using an efficient Hamiltonian simulation subroutine acts on a block-encoded Hamiltonian. We show that the resulting algorithm can be used to solve quantum phase estimation (QPE) or quantum energy estimation (QEE) {with competitive complexity scaling.} 

We then use recent results for block-encoding implementations to derive the Clifford + T complexity bound for QPE with respect to model-relevant parameters of the Hamiltonian and the desired precision. With this result, we demonstrate an efficient algorithm for QEE beginning from any linear combinations of Pauli strings. 

In this way, we argue that a well-understood and long-standing idea retains practical legitimacy for fault-tolerant era algorithms, once the costs of Hamiltonian simulation are accounted for.

\end{abstract}

\section{Introduction}
Quantum phase estimation and quantum energy estimation algorithms (QPE and QEE) have been steadily developed in the circuit model since the 1990s \cite{kitaev_quantum_1995, Nielsen_Chuang_2010}. However, QEE-like problems were already considered in von Neumann's measurement procedure \cite{childs_quantum_2002, von_neumann_mathematical_2018}. QEE with von Neumann's method has fallen largely out of focus, with newer methods developed for digital quantum devices becoming more popular \cite{gilyen_quantum_2019, Lin2020nearoptimalground, donggep2022, rall_faster_2021}. 

This work does not argue that von-Neumann's measurement protocol based QEE (vN-QEE) can supersede  existing methods. Instead, we show that it can be made competitive, meaning that  {the asymptotic query complexity  is efficient for vN-QEE.}

We then show how to efficiently implement QEE for any Hamiltonian given as a linear combination of Pauli strings. In doing so, we highlight that a simple, pedagogically accessible, and physically motivated QEE can still give efficient results for a large and highly relevant class of systems.

In the QPE problem, one assumes access to a unitary $U=e^{iH}$ and a state preparation scheme for the eigenstate $\ket{\lambda}$ of $U$, corresponding to the eigenvalue $e^{i\lambda}$. QPE is the problem of finding a  {$\epsilon$-accurate} estimate of $\lambda$,  succeeding with success probability at least $1-\delta$ for $\delta\in(0, 1)$. QPE is closely related to QEE, the problem of finding the eigenstates of the generator $H$. In QEE, $H$ is typically accessed through a unitary block-encoding $BE(H)$.

Block-encodings have quickly become a popular strategy for encoding data \cite{Childs_2009, camps_explicit_2023}; they are convenient to manipulate using a family of templates for quantum computing: quantum signal processing (QSP),  and quantum eigenvalue transformation (QET) \cite{low_hamiltonian_2019, low_method_2016}, and quantum singular value transformation (QSVT) \cite{gilyen_quantum_2019}. Herein, we will use "QSP+"  to refer to QSP/QET/QSVT, whenever the details of the template do not matter.

We show that in combination with efficient Hamiltonian evolution implemented in QSP+, vN-QEE can be a simple, efficient algorithm whenever a bound on the gap between the eigenstate of interest and others is known. Key to this statement is demonstrating that the coupling of the pointer and system of interest introduces only a logarithmic complexity overhead. Then, QSP+ techniques can be used to efficiently encode an approximation of the eigenstate in the pointer system.  {We then use a result from \cite{gilyen_quantum_2019} to derive a QPE routine as a corollary of our main theorem.} 

 {It is expected that system-specific parameters will determine the costs of a QEE or QPE algorithm, namely the system dimension $2^n$, $n\in\mathbb{I}_{\geq 0}$, the spectral gap $\Delta_0$, and the spectral norm $\alpha$. In particular, to resolve eigenstates QEE typically requires $\epsilon_{vN}<|\lambda_k-\lambda_{j\neq k}|$ and so expected to be $\alpha, n$ dependent. The exact relationship between $n, \Delta_0, \alpha$ has been studies  both in asymptotic limits \cite{khorunzhy2004spectralnormlargeband,Fronk2023NormCR} and with quantitative studies \cite{gergelitsnumerical2022,becker2026approximatingoperatornormlocal, Loaiza_2023}. }

Our QEE algorithm scales optimally, with $\mathcal{O}(\frac{\alpha}{\epsilon}\log_2(\delta^{-1}))$ in \cref{thrm:qeequeryoptimal}. However, \cref{cor:qpequery} demonstrates that our algorithm is logarithmically sub-optimal in the QPE case \cite{mandetight2023}.  {A fuller comparison can be found in \cref{sec:relatedwork} and \cref{tab:qpecompquery}.}

In \cref{sec:prelim}, we review basic facts about block-encodings and QSP+. Likewise, in \cref{sec:vonneumann} we review the von Neumann measurement protocol and show how either the pointer system or the time-evolution can be used to mediate QEE precision. In \cref{sec:qpealgs}, we give the main results of this work: a QEE algorithm using QSP+ Hamiltonian simulation and complexity  {bounds. In all settings, creating oracle access to $H\otimes P$ introduces only a small overhead.}

\Cref{thrm:QEE-LPC} is the result of the most practical interest: we give Clifford plus $T$ gate complexities and depths for QEE, for any Hamiltonian given as a linear combination of Pauli strings.  The depth and ancilla requirements of such a preparation are well understood from \cite{Zhang_2024}. The query complexity of \cref{thrm:QEE-LPC} is stated with respect to only the accuracy of the QEE and physically meaningful Hamiltonian parameters. Although this is really a synthesis of existing block-encoding complexities, Hamiltonian simulation, and QPE protocols, our argument demonstrates that a large and highly relevant class of Hamiltonians admit efficient QEE and QPE algorithms.

\subsection{Related work}\label{sec:relatedwork}
Quantum phase estimation algorithms almost all follow the same high-level strategy\textemdash encode an estimate of $\lambda$ on $r$ ancillary qubits using controlled powers of $U$ applied on a superposition of computational basis states in the circuit model (more generally, a superposition of position basis states).

In recent years, strong techniques for problems related to either QEE and QPE have been developed \cite{ge2018fastergroundstatepreparation, king2025quantumsimulationsumofsquaresspectral, Lin2020nearoptimalground, rall_faster_2021, donggep2022, chakrabortypowers2019}. Many also apply to the standard QEE/QPE so we will discuss them herein.

We summarize the relevant literature and our algorithms in \cref{tab:qpecompquery}. To simplify the exposition, we will assume that all sub-normalizations are equal to the spectral norm.  The most important distinction is that we assume that the overlap between the initial state and the desired eigenstate, $\braket{\lambda|\psi_0}\geq \gamma$, is exponentially close to one. More specifically, we assume $\gamma=1-\mathcal{O}(2^{-n})$. Our algorithm would require $\mathcal{O}(\gamma^{-2})$ trials to succeed with a smaller $\gamma$, from the same argument given to bound the necessary number of trials with standard QPE \cite{ge2018fastergroundstatepreparation, linheisenburg2022}. 

Finally we will  {focus on coherent algorithms in this work}.  {In a coherent algorithm, } the entire protocol runs on a quantum device and then one measurement of $r$ qubits occurs at the end of the computation. Crucially, the definition only applies when $\gamma\rightarrow 1$ and ignores the costs of running many trials to increase the success probability.  { Kitaev's phase estimation \cite{kitaev_quantum_1995} or Rall's QSVT-based algorithms \cite{rall_faster_2021} can both be implemented non-coherently with $\mathcal{O}(1)$ qubits. Additionally, \cite{Lin2020nearoptimalground} can be used for ground state problems.}

Leading QPE algorithms use QSP+ subroutines like spectral amplification\cite{gilyen_quantum_2019, chakrabortypowers2019}, eigenstate filtering \cite{Lin2020nearoptimalground}, or prepare a specialized oracle that allows for a Kitaev-like algorithm \cite{rall_faster_2021}. {Because good expositions of the algorithms in \cref{tab:qpecompquery} can be found elsewhere (especially \cite{Lin2020nearoptimalground, king2025quantumsimulationsumofsquaresspectral, linheisenburg2022}), we do not discuss each algorithm in depth.}

\begin{sidewaystable}
    \centering
    \resizebox{25cm}{!}{
    \begin{tabular}{|c|c|c|c|c|}\hline
        Algorithm &   $\cost{U}$ (QPE) & or $\cost{BE(H)}$ (QEE) & r  & caveats\\ \hline\hline
         QPE lowerbound \cite{mandetight2023} & $\Omega\left(\frac{1}{\epsilon}\log(1/\delta)\right)$ & N/A & $\cost{\log(1/\epsilon)}$ & assume $\gamma=1$ \\
         QSVE \cite{gilyen_quantum_2019, chakrabortypowers2019} & N/A &  $\mathcal{O}\left(\frac{\alpha}{\epsilon}\text{polylog}\left(\frac{1}{\delta}\right)\right)$ & $\log(1/\epsilon)$ & $\Delta\leq 1$, $\gamma=1$ \\
         Ge et al. \cite{ge2018fastergroundstatepreparation} & $\tilde{\mathcal{O}}\left(\frac{1}{\gamma \epsilon^{3/2}}\right)$ & N/A & $\mathcal{O}(\log(\epsilon^{-1}))$ &  $\epsilon=\tilde{\mathcal{O}}(\Delta)$\\
        QSVT coherent \cite{rall_faster_2021} & $\cost{\frac{1}{\epsilon}\eta^{-1}\log(1/\delta)}$ & $\cost{\eta^{-1}\log(\epsilon^{-1})\left(\frac{1}{\epsilon}+\log(\eta^{-1})\right)}$ & $\cost{\log(1/\epsilon)}$ & $\opnorm{H}\leq 1$, $\lambda_j\neq \left[\frac{x}{2^r}, \frac{x}{2^r}+\eta\right]$, $\gamma=1$\\
          \cref{thrm:qeequeryoptimal}, \cref{cor:qpequery} & $\tilde{\mathcal{O}}\left(\frac{\alpha}{\epsilon}\log(\frac{1}{\delta})\right)$ & ${\mathcal{O}}\left(\frac{\alpha}{\epsilon}\log(\frac{1}{\delta})\right)$ & $\cost{\log(1/\epsilon)}$ & $\gamma$ high, $\epsilon\leq \Delta_k$\\ \hline

    \end{tabular}
    }
    \caption{QPE and/or QEE algorithms, obtaining an $\epsilon$-close estimate of $\lambda$. Each algorithm succeeds with success probability $1-\delta$ given overlap $\gamma$, which is restricted in some algorithms to be high. We assume that $\Delta_k$ is known, and also $\alpha$.  Note that Kitaev's phase estimation with median amplification scales with the lower bound \cite{mandetight2023}. Some algorithms also have stipulations on the eigenvalue domain, for which we use $\eta\in\mathbb{R}_{+}$. We use $\tilde{\mathcal{O}}$ to denote the suppression of logarithmic factors of $\alpha, \log(\delta^{-1}), \epsilon$.}
    \label{tab:qpecompquery}
\end{sidewaystable}

We compare the algorithms for $\gamma\rightarrow 1$, meaning for the standard QPE or QEE problem. More generally, our algorithm will have subpar $\gamma$ scaling, and will not be competitive with \cite{donggep2022, king2025quantumsimulationsumofsquaresspectral, ge2018fastergroundstatepreparation}.

 {\Cref{thrm:qeequeryoptimal} has competitive scaling with QSVE} \cite{gilyen_quantum_2019, chakrabortypowers2019} and QSVT coherent \cite{rall_faster_2021}. It is unclear at an asymptotic level which algorithm will better succeed for some given problems, as  $\alpha, \eta$ are not fully equivalent criteria. However, compared to the lower bound \cite{mandetight2023}, and the QSVT algorithm proposed in \cite{rall_faster_2021}, vN-QPE (\cref{cor:qpequery}) is logarithmically  weaker. If we relax the assumption $\gamma\approx 1$, then \cite{ge2018fastergroundstatepreparation} will still have the best complexity scaling with respect to the overlap.

Finally, note that one could derive QPE algorithms from QEE algorithms using \cref{lemma:gilyenBEoracle} from \cite{gilyen_quantum_2019}, similarly to our  {\cref{cor:qpequery}}. However, this would not change the comparison in this section.
\section{Preliminaries}\label{sec:prelim}
Encoding Hamiltonians in unitary matrices with a larger rank has become standard in quantum algorithms.  In QSP+, one encodes the input Hamiltonian and then creates the \textit{signal operator}, an oracle which is called in each step of the QSP+ protocol. 

We will use a slight generalization from the standard definition of a block-encoding\footnote{Our definition would most properly be called an approximate standard-form-encoding but to remain compatible with existing literature, we will always use the terms "block-encoding" and "qubitized block-encoding" in the main text.}
\begin{definition}[Block-encoding, generalized from \cite{gilyen_quantum_2019, low_hamiltonian_2019}]\label{def:blockencoding}
   Given $n_{anc}, s\in\mathbb{Z}_+$ and Hamiltonian $H$, assume access to some unitary $BE(H): \mathcal{H}_{n_{anc}}\otimes \mathcal{H}_s\rightarrow \mathcal{H}_{n_{anc}}\otimes \mathcal{H}_s$, a state preparation scheme such that $G\ket{0}_{n_{anc}}=\ket{G}\in \mathcal{H}_{n_{anc}}$, and  $\epsilon_{BE}\geq 0$, $\beta\in\mathbb{R}_+$. Then, $BE(H)$ is a $\left(\beta, n_{anc}, \epsilon_{BE}\right)$ block-encoding of $H$ if 
   \begin{equation}
       \opnorm{\beta\bra{G}BE(H)\ket{G}-H}\leq  {\epsilon_{BE}}.
   \end{equation}
\end{definition}
By $\bra{G}BE(H)\ket{G}$, we mean a projection to $\ket{G}$ and then a partial trace over the $n_{anc}$ ancillary qubits, that is $\text{Tr}_{n_{anc}}\left[I_{s} \ket{G}\bra{G}\otimes I_{s}BE(H)\right]$.  Below, we will always assume that $\beta\geq \opnorm{H}$.

In QET, one needs to use block-encodings with a particular property, namely the qubitized standard-form encoding introduced in \cite{low_hamiltonian_2019} whose construction is given in \cref{sec:qubitizedblockencodings}. A key feature of qubitized block-encodings is that they have eigenvalues of the form\footnote{Note that we have fixed a global phase to get to this definition of the eigenvalues, see \cref{sec:qetoracles}.} $e^{\pm i\arcsin(\lambda/\beta)},$  where $\lambda$ are the eigenvalues of $H$ \cite{low_hamiltonian_2019}.

We will consider two ways to create a block-encoding for a given Hamiltonian: either access to the  {Pauli} decomposition of the Hamiltonian or the ability to prepare the unitary oracle $e^{iH}$.

One can efficiently prepare a block-encoding of $H$ from $U$, using a protocol from \cite{gilyen_quantum_2019}, reproduced in \cref{lemma:gilyenBEoracle}. This will allow for the QEE protocol defined in \cref{alg:QSPQPE} to also solve the QPE problem.

We also consider QEE bounds starting from one of the most common representations of time-independent Hamiltonians, the linear combination of Pauli strings (LCP)
\begin{equation}
    H=\sum_{l}\alpha_lP_l.
\end{equation}
Above, $P_l$ is any element in the Pauli basis for $U(n)$, there are a total of $\left|P\right|$ terms, and $\alpha=\sum_l|\alpha_l|$. Note that $\alpha$ bounds the norm of $H$.  We denote the operator norm of the matrix with  $\opnorm{H}$, which for Hermitian matrices has the simpler forms such as the spectral norm.

Every fermionic quantum system has such a decomposition; spin systems are often represented in this form, and there exist techniques to find the Pauli decomposition of Hermitian matrices \cite{jones2024decomposingdensematricesdense, Hantzko_2024, Vidal_Romero_2023}. In many quantum simulation settings, the the number of Pauli terms required to encode $H$ directly dictates how difficult the system will be to simulate. So, there are strong physical motivations to work with LCP's.

From \cite{Zhang_2024}, one can encode a LCP in a block-encoding with an efficient cost in Clifford and $T$ gates. Clifford and $T$ gates are the set of the generators of the Clifford group along with the $T$ gate; We denote it $\mathcal{G}_{CpT}$. $\mathcal{G}_{CpT}$ is approximately universal in the sense of the Solovey-Kitaev theorem \cite{Nielsen_Chuang_2010} and reasonably common, especially for resource theories \cite{howardapplication2017, Veitch_2014, gottesman1998heisenbergrepresentationquantumcomputers} and fault-tolerant implementations \cite{zhoumethadology2000, bravyiuniversal2005}. 

\subsection{QSP+}
The task of QSP is to perform the following action on some unitary 
\begin{align}
    U&=\sum e^{\pm i\arccos(\lambda)} \ket{\lambda}\bra{\lambda}\rightarrow \nonumber\\
    \bra{a_{QSP}}&U_{QSP}\ket{a_{QSP}}=P(U)\\
    &=\sum P(e^{\pm i\arccos(\lambda)}) \ket{\lambda}\bra{\lambda}.
\end{align}
Where $\ket{a_{QSP}}$ is a basis measurement on the QSP rotation qubit, usually selected as $\ket{+}$ or $\ket{0}$.

The task of QET is to apply $P$ to the eigenvalues of a Hermitian operator accessed within a qubitized block-encoding, that is
\begin{align}
    H&=\bra{G'}SBE(H)\ket{G'}\rightarrow\\
     P(H)&=\bra{G',a_{QET}}U_{QET}\ket{G',a_{QET}}.
\end{align}
where $\ket{G'}\bra{G'}$ is the projector which selects $H$ within the qubitized block-encoding.

QSP+ circuit and theorems are well-known elsewhere for various conditions on the polynomial $P$ \cite{motlagh_generalized_2023, gilyen_quantum_2019,low_hamiltonian_2019, low_method_2016, Haah2019product, martyn_grand_nodate}, we review QET in more detail in \cref{sec:qetoracles}. We consider QSP+ with Laurent polynomials with total degree $d$, obeying the following 
\begin{enumerate}
 \item $P:  \mathbb{C}\rightarrow \mathbb{C}, \quad \forall z\in U(1), \quad  \linfnorm{P}\leq 1$
    \item $\text{Re}(P)(z)=\pm \text{Re}(P)(z^{-1})$ and $\text{Im}(P)(z)=\pm \text{Im}(P)(z^{-1})$ for all $z\in U(1)$
    \item $P(z)$ is $\epsilon_{approx}$-close in the $l_\infty$-norm to some function $f$.
\end{enumerate}
These polynomials fulfill the conditions for QSP and QET as are found in \cite{low_hamiltonian_2019, Haah2019product}. 
With Laurent polynomials of degree $d$, the QET circuit requires $2d$ controlled operations of the {signal operator} and $2d+1$ single-qubit rotation gates, the \textit{signal processing operators}. 

Say that the initial state has some decomposition in the eigenbasis of the Hamiltonian, $\sum c_j\ket{\lambda_j}\otimes \ket{G'}$, $\sum_j c_j=1$. For such initial states, the QET circuit prepares
\begin{equation}
  \sum_jc_j{P'\left(\frac{\lambda_j}{\alpha}\right)}\ket{\lambda_j}\ket{G',a_{QET}}+\ket{\psi_{QET}^{\perp}}.
\end{equation}
The probability of obtaining the desired branch of the state upon measurement is 
 {
\begin{equation}
\left|{\sum_jc_j{P'\left(\frac{\lambda_j}{\alpha}\right)}}\right|^{2}\label{eq:qspsuccprob}
\end{equation}
}

\section{von Neumann measurement scheme}\label{sec:vonneumann}
In von Neumann's measurement scheme, the system Hamiltonian is coupled to a free particle with momentum $p$, so the full Hamiltonian is 
\begin{equation}
    H(s)\otimes \vec{p}.
\end{equation}
Once one assumes that the particle mass is large enough to neglect kinetic energy (and sets $\hbar=1$), the evolution is $e^{-itH(s)\otimes p}$. Assuming the free particle starts off very localized in space, initially $\ket{x}$ is a narrow Gaussian wave packet centered at $x=0$. Because the simulation algorithm used herein assumes a time-independent Hamiltonian, we immediately simplify the notation to $H(s)=H, \lambda(s)=\lambda$.

The evolution starting from the $k$th eigenstate $\ket{{\lambda}_k, x=0}$ evolves to
\begin{equation}
    \ket{{\lambda}_k,x=t\lambda_k}
\end{equation}
All one has to do to obtain $\lambda_k$ after this evolution is measure the pointer state. This is already von Neumann's measurement prescription for obtaining  {$i \ln(U)$}; reference  \cite{childs_quantum_2002} makes this connection more explicit and presents an unstructured search algorithm using the procedure.

Implementing time evolution of $ e^{-itH\otimes p}$ with QET in a digital/discrete system requires one to discretize the continuous variable $\Vec{x}$, and the momentum operator becomes  {either}
\begin{equation}
   \vec{\tilde{p}}=\sum_{j=1}^{r}2^{-j}\frac{1-\sigma_z^{(j)}}{2} \Rightarrow \vec{\tilde{p}}\ket{z}=\frac{z}{2^r}\ket{z}\label{eq:discretemomentumop}.
\end{equation}
 {or}
 {\begin{align}
     \vec{{p}}&=\sum_{j=0}^{{r}-1}2^j\frac{I-\sigma_Z^{(j)}}{2}, \Rightarrow {p}\ket{z}=z\ket{z}\label{eq:largenormmomentum}.
\end{align}}

In  {both \cref{eq:discretemomentumop} and \cref{eq:largenormmomentum}}, the momentum operator basis is fixed to be the computational basis states over a $2^r$ dimensional space, $\{\ket{z}\}$ for $z\in \mathbb{Z}_+$ where $z< 2^r$ for some integer $r$.  {\Cref{eq:largenormmomentum} corresponds to the standard register size used in QPE, and \cref{eq:discretemomentumop} is a register used in \cite{childs_quantum_2002} to create a quantum search algorithm, which is expected to work with  $r=\mathcal{O}(1)$ register. Herein, we consider only \cref{eq:largenormmomentum}.}

 {The system initial state is
\begin{equation}
    \ket{\lambda_{k}}\ket{x(0)}=\frac{1}{2^{{r}/2}}\sum_{z=0}^{2^{{r}}-1}\ket{\lambda_k, z}.
\end{equation}
}
Beginning from $\ket{\lambda_{k}}\ket{x(0)}$, after the evolution the state is 
\begin{equation}
   {\ket{\lambda_k}\otimes \ket{x(0)}\rightarrow\frac{1}{2^{r/2}}\ket{\lambda_k}\sum_{z=0}^{2^r-1} e^{-i{t\lambda_kz}}\ket{z}.}\label{eq:eigstatepostHS}
\end{equation}

Finally the inverse quantum Fourier transform (iQFT) \cite{Nielsen_Chuang_2010} on the $r$ register results in state
 {
\begin{align}
  iQFT \ket{x({t}), \lambda}&=\frac{1}{2^{{r}}}\sum_{z=0}^{2^{{r}}-1}\sum_{x=0}^{2^{{r}}-1} e^{i\frac{2\pi x z}{2^{{r}}}}e^{-i{t}\lambda z}\ket{z, \lambda}\label{eq:vNmfinalstate}
\end{align}
}
where the probability amplitude of any pointer computational basis state is
 {
\begin{align}
    \kappa_x=\frac{1}{2^{{r}}}\sum_{z=0}^{2^{{r}}-1}e^{iz\left(\frac{2\pi x}{2^{{r}}}-{t}\lambda_k\right)}.\label{eq:QEEperfectprobamp}
\end{align}
}
This is already a similar state to the result of textbook QPE, and so we can replicate the usual analysis of \cite{Nielsen_Chuang_2010, Brassard_2002}. Expanding \cref{eq:QEEperfectprobamp} as a complex geometric series, one arrives at the usual expression, where $n\in\mathbb{Z}_{\geq 0}$ and  {$x\neq \frac{2^{r}\lambda t}{2\pi}$}
 {
\begin{align}
    \left|k_{x}\right|^2&=
        \frac{\sin^2\left({\pi\left(x-2^r\frac{\lambda t}{2\pi}\right)}\right)}{4^r\sin^2\left(\frac{\pi}{2^r}\left(x-2^r\frac{\lambda t}{2\pi}\right)\right)}\label{eq:childssuccprop}
\end{align}
When  $x$ is measured with certainty, then $\lambda=\frac{2\pi x}{2^rt}$.} In the phase estimation problem, usually only one eigenvalue $\lambda_k$ is of interest. 

Beginning from \cref{eq:childssuccprop}, one can derive the result using similar methods to \cite{Brassard_2002}. We get
\begin{equation}
    |\kappa_x|^2\geq 1-\frac{1}{2(k-1)}, \quad k>1\label{eq:kappaboundbrassard}
\end{equation}
where $|\kappa_x|^2$ is the probability that answer $\tilde{x}/2^r$ is such that 
 {
\begin{equation}
    \left|\frac{2\pi\tilde{x}}{t2^r}-\lambda\right|\leq \frac{k}{2^r}.
\end{equation}
}
For $\kappa_x$  to have success probability higher than $1-\delta$ for some $\delta\in[0, 1)$, then $k$ can be restricted by $\delta$
\begin{align}
    k&=\lceil\frac{1}{2\delta}\rceil+1.\label{eq:kconstraint}
\end{align}

Denote the distance  {(gap)} between any two eigenvalues as 
\begin{align}
    \Delta_{ij}=\left|\lambda_j-\lambda_i\right|,
\end{align}
and then fix $\Delta_k=\max_{l}\Delta_{kl}$. If one is interested in measuring the ground state energy, then $\Delta_0=\Delta$ is the spectral gap. 

With a system prepared in \cref{eq:vNmfinalstate}, the $r$-register estimate, $\tilde{x}$,  defines the phase estimate 
 {
\begin{equation}
    \tilde{\lambda}=\frac{2\pi\tilde{x}}{2^rt}.
\end{equation}
} 
The QEE will only succeed if (i) the probability of measuring another bit string $\tilde{x}'\neq \tilde{x}$ is always small, which is managed by \cref{eq:kconstraint}, and  (ii) the pointer system is large enough to encode an answer to some desired accuracy $\epsilon_{vN}$. 

With $\left|{x}-{\tilde{x}}\right|\leq {k}$, 
 {
\begin{align}
  \left|\lambda-\tilde{\lambda}\right|&= \frac{2\pi}{2^{{r}} {t}}\left|x-\tilde{x}\right|\leq  {\frac{2\pi k}{2^{{r}}{t}}}\label{eq:epsilonvN}.
\end{align}
}

 {
The algorithm cannot successfully distinguish between eigenvalues unless the required precision is less than the gap, i.e. $ \left|\lambda-\tilde{\lambda}\right|\leq \epsilon_{vN}\leq \Delta_k$.

 {The bit precision is $k/2^r\leq \epsilon_{bit}\in(0, 1)$, with 
\begin{equation}
    r\geq \log_2\left(\frac{k}{\epsilon_{bit}}\right).
\end{equation}
leading to 
\begin{equation}
    r\geq \log\left(\frac{1}{\epsilon_{vN}}\right)\label{eq:rconstraintoptimalquery},
\end{equation}
matching the register used in the standard QPE problem.
}

The resulting choice of $t$ is
\begin{align}
     {{t}} &\geq 2\pi \label{eq:tauconstraintoptimalquery},\\
    \Rightarrow   \left|\lambda-\tilde{\lambda}\right|&\leq \frac{k}{2^{{r}}}.
\end{align}

\subsection{Hamiltonian Simulation approximation}
The polynomial approximation and bounds for QET Hamiltonian simulation were developed in \cite{low_hamiltonian_2019, gilyen_quantum_2019} by truncating the Jacobi-Anger expansion. In coordinate $z\in U(1)$, 
\begin{align}\label{eq:laurent_HSpoly}
    P_{HS}&=\mathcal{A}(z)-i\mathcal{B}(z),\\
    \mathcal{A}(z)&=J_0(t)+\sum_{k=-R}^{R}(-1)^k{J_{2k}(t)}z^{2k},\\
    \mathcal{B}(z)&=\sum_{-R-1}^R(-1)^k{J_{2k+1}(t)}z^{2k+1}.
\end{align}
where the total degree is $d=2R+1$. To obtain an $\epsilon_{QET}$-close approximation, one can use the following bound

 \begin{equation}\label{eq:rupperbound}
        d=\mathcal{O}\left(t\log\left(\frac{1}{\epsilon_{QET}}\right)\right).
\end{equation}
 {$\epsilon_{QET}$ is a bound on the error introduced by using $P_{HS}$, an imperfect estimate of the target function in the QET routine.}

\subsection{QPE with imperfect subroutines}
So far, we have assumed that $e^{it H\otimes p}$, $\ket{\lambda}$, and the QFT's can be perfectly implemented. However, these assumptions can all be relaxed. 

Say one implements a Hamiltonian simulation algorithm to precision $\epsilon_{HS}\in[0, 1)$. Beginning from some $\ket{\lambda_k, z}$, we have 
\begin{align}
    \left|\left|U_{QSP}\ket{\lambda_k, z}-e^{-it H}\ket{\lambda_k, z}\right|\right|\leq \epsilon_{QET}
\end{align}
After the iQFT is applied to $U_{QSP}\ket{\lambda_k, z}$, one obtains 
\begin{align}
\left|\left|\ket{\lambda_k}\sum_{x=0}^{2^r-1} \kappa_{x}\ket{{x}}-iQFT\cdot U_{QSP}\ket{\lambda_k, z}\right|\right|\leq \epsilon_{QET}\label{eq:tonorm}
\end{align}
 So then, the norm of $iQFT\cdot U_{QSP}\ket{\lambda_k, z}$ is at most $1+\epsilon_{HS}$, leading to a normalized state 
\begin{align}
   \frac{iQFT\cdot U_{QSP}\ket{\lambda_k, z}}{\sqrt{1+\epsilon_{QET}}}
\end{align}
instead of \cref{eq:vNmfinalstate}.

The success probability becomes
 {
\begin{equation}
    \left|\tilde{\kappa}_x(\epsilon_{HS})\right|^2\geq\frac{ \left|{\kappa}_x\right|^2}{(1+\epsilon_{HS})}.
\end{equation}
}
We also consider imperfect eigenstate preparation $\ket{\psi_0}=\sum c_j\ket{\lambda_j}$, where $c_{j\neq k}$ will be assumed to be small. Measuring in the Hamiltonian eigenbasis, the success probability of obtaining $\lambda_k$ is $1-\sum_{j\neq k}c_j^2$. Using  $\ket{\psi_0}$ instead of $\ket{\lambda_k}$ as the initial state in QPE, one must account for the possibility of measuring a result for eigenstate $\lambda_j\neq \lambda_k$. By linearity, 
\begin{align}
    \left|\tilde{\kappa}_x(\epsilon_{HS}, \{c_j\})\right|^2& {\geq \frac{{1-\sum_{j\neq k}c_j^2}}{(1+\epsilon_{HS})}|\kappa_x|^2\label{eq:succprobvnapproxhs}.}
\end{align}

\section{vN-QPE algorithms}\label{sec:qpealgs}

 {We state \cref{thrm:qeequeryoptimal} showing that the QEE problem can be solved with \cref{alg:QSPQPE}, and then in \cref{cor:qpequery} we show that QPE can be solved with near-optimal query complexity with the same strategy.}
%\begin{algorithm}
%\caption{vN-QPE with QET. $W(H)$ is a qubitized $(\beta, n_{anc},  {\epsilon_{vN}}, \ket{G'})$ block-encoding of $H$, where $H$ has eigenstate $\ket{\lambda_k}$. We assume three registers with respectively $[r], [n_{anc}+\lceil\log_2\text{rank}(H)\rceil]$, $[1]$ qubits, and $\vec{p}$ defined  in  \protect\cref{eq:largenormmomentum}.}
%\label{alg:QSPQPE}
%\begin{algorithmic}[1]
%\Require {Eigenstate $\ket{\lambda_k}$ of $H$, $r\in \mathbb{Z}_{\geq 1}$, and unitary $W(H)$ where $\beta \bra{G'}W(H\otimes \Vec{p}')\ket{G'}=H$ for some $\beta\in\mathbb{R}_{> 0}$, $t\in\mathbb{R}_{\geq 0}$}
%\State Prepare $\ket{+, G', \lambda_k, 0^{\otimes r}}$
%\State Apply Hadamard-Walsh transform on $[r]$  
%\State Run QET Hamiltonian simulation  on $W(H\otimes \Vec{p})$ for time $\beta t$
%\State Run iQFT on $[r]$
%\State Post-select on $[1, n_{anc}]$ being in $\ket{+, G'}$
%\State Measure $[r]$ register in the computational basis, obtaining $\tilde{x}$. Compute estimate $\tilde{\lambda}$ using $\tilde{\lambda}=\frac{2\pi}{2^r t}\tilde{x}$
%\end{algorithmic}
%\end{algorithm} 

\begin{algorithm}
\caption{vN-QPE with QET. $W(H)$ is a qubitized $(\beta, n_{anc},  {\epsilon_{vN}}, \ket{G'})$ block-encoding of $H$, where $H$ has eigenstate $\ket{\lambda_k}$. We assume three registers with respectively $[r], [n_{anc}+\lceil\log_2\text{rank}(H)\rceil]$, $[1]$ qubits, and $\vec{p}$ defined  in  \protect\cref{eq:largenormmomentum}.}
\label{alg:QSPQPE}
\KwIn {Eigenstate $\ket{\lambda_k}$ of $H$, $r\in \mathbb{Z}_{\geq 1}$, and unitary $W(H)$ where $\beta \bra{G'}W(H\otimes \Vec{p}')\ket{G'}=H$ for some $\beta\in\mathbb{R}_{> 0}$, $t\in\mathbb{R}_{\geq 0}$}
State Prepare $\ket{+, G', \lambda_k, 0^{\otimes r}}$\;
Apply Hadamard-Walsh transform on $[r]$\;
Run QET Hamiltonian simulation  on $W(H\otimes \Vec{p})$ for time $\beta t$\;
Run iQFT on $[r]$\;
Post-select on $[1, n_{anc}]$ being in $\ket{+, G'}$\;
Measure $[r]$ register in the computational basis, obtaining $\tilde{x}$. Compute estimate $\tilde{\lambda}$ using $\tilde{\lambda}=\frac{2\pi}{2^r t}\tilde{x}$\;
\end{algorithm}

We will state the results in terms of \textit{any} $1, 2$ qubit gates, and in the next we will take advantage of the clearer complexity and depth statements available through \cref{thrm:zhang7}.

\subsection{QEE with near-optimal complexity}
\begin{theorem}[query-focused quantum energy estimation]\label{thrm:qeequeryoptimal}
    Consider a rank $2^n$ Hamiltonian $H$, such that its $k$th eigenvalue has gap $\Delta_k$. Assume access to a $\left(\beta, n_{anc}, \epsilon'\right)$ block-encoding of $H$, where $\epsilon'\leq \frac{\epsilon_{vn}^2\delta}{\beta\log(\delta^{-1})}$, and an approximation of the eigenstate corresponding to $\lambda_k$, $\ket{\psi_0}=\sum_jc_j\ket{\lambda_j}$. We say that for $j\in [2^n]$, $c_j^2\leq \frac{\delta}{3\cdot2^{n-1}}$ for all $j\neq k$ and some $\delta\in (0, 1)$.
    Then, \cref{alg:QSPQPE} prepares an $\epsilon_{vN}\leq \Delta_k$-close approximation of $\lambda_k$ with success probability at least $1-\delta$, using  $\log(1/\epsilon_{vN})+n+n_{anc}+2+\log\log(\epsilon_{vN}^{-1})$ qubits.
    The protocol requires 
    \begin{align}
       \mathcal{O}\left(\frac{\beta}{\epsilon_{vN}}\log_2\left(\frac{6}{\delta}\right)\right)\nonumber
    \end{align}
calls to the controlled block-encoding and its inverse, and 
\begin{align}
    {\mathcal{O}}&\left(\log_2^2\left(\frac{1}{\epsilon_{vN}}\right)+\frac{\beta}{\epsilon_{vN}}\log_2\left(\frac{1}{\delta}\right)\right.\nonumber\\
    &\cdot \log_2\left(\frac{1}{\epsilon_{vN}}\right)\log\left(\frac{\beta^2\log_2\left(\frac{1}{\delta}\right)}{\delta\epsilon_{vN}}\right)\nonumber
\end{align}
additional $1, 2$ qubit gates.
\end{theorem}

\begin{proof}
   First,  $\opnorm{{p}}=2^{{r}}=\frac{1}{\epsilon_{vN}}$ and we use \cref{thrm:zhang7} to prepare a $\left(\frac{1}{\epsilon_{vN}}, \log\left[\log_2\left(\frac{1}{\epsilon_{vN}}\right)\right], \frac{\epsilon_{BE}}{\beta}\right)$ block-encoding of $ {p}$.  \Cref{lemma:betensorlemma} then obtains a $\left(\beta/\epsilon_{vN}, n_{anc}+\log\left[\log_2\left(\frac{1}{\epsilon_{vN}}\right)\right], \epsilon_{BE}\right)$ block-encoding for $H\otimes {p}$.

  Now, one solves the Hamiltonian simulation problem for time $t'$, where $t'=\beta {t}$, and precision $\epsilon_{QSP}$.  The costs of doing so scales with polynomial degree $d=\mathcal{O}(t'\log(\epsilon_{QSP}^{-1}))$  from \cref{eq:rconstraintoptimalquery}. 

    After post-selection to the $\ket{+,G'}\bra{+,G'}$ subspace containing the desired state, the operator norm error between the ideal circuit and the $\epsilon_R, \epsilon_{BE}, \epsilon_{QSP}$ close approximation is, using \cref{lemma:approxqsperrorbound}, 
\begin{align}
    &\opnorm{e^{it H}-\bra{+G'}\Tilde{U}_{QSP}\ket{+G'}}\nonumber\\
    &\leq \epsilon_{QSP}+2d\epsilon_{BE}\\
    &:= \epsilon_{HS}.\label{eq:opnormqspwerror}
\end{align}
We arrive at the last line by restricting $\epsilon_{BE}\leq \epsilon_{HS}/2d$ and $\epsilon_{QSP}\leq \epsilon_{HS}/2$

From \cref{eq:succprobvnapproxhs}, the success probability is 
\begin{align}
      \left|{\kappa}_x'(\epsilon_{HS}, \{c_j\})\right|^2&\geq\frac{{1-\sum_{j\neq k}c_j^2}}{(1+\epsilon_{HS})}\frac{1}{4^{{r}}}\nonumber\\
      &\cdot\left(\sum_{z=0}^{2^{{r}}-1}e^{iz\left(\frac{2\pi x}{2^{{r}}}-{t}\lambda_k\right)}\right)^2
\end{align}
It is sufficient to restrict $\epsilon_{HS}=\frac{{\delta}}{{3}}$, $c_j^2\leq \frac{{\delta}}{3\cdot 2^{(n-1)}}$, and then
\begin{align}
    \left(1-\frac{1}{2(k-1)}\right)&\geq \left(1-\frac{\delta}{3}\right)\\
    \Rightarrow& k=\lceil\frac{3}{2\delta}\rceil+1.
\end{align}
The success probability becomes
\begin{align}
     \left|{\kappa}'_x(\epsilon_{HS}, \{c_j\})\right|^2&\geq \frac{\left(1-\frac{\delta}{3}\right)^2}{\left(1+\frac{\delta}{3}\right)}\\
     &\geq \left(1-\frac{\delta}{3}\right)^3\\
     &\geq \left(1-\delta\right).
\end{align}
    Then, one simply works out 

    \begin{align}
        \epsilon_{QET}&\leq \frac{\delta}{6}\\
    d&=\mathcal{O}\left(\frac{\beta}{\epsilon_{vN}}\log_2\left(\frac{6}{\delta}\right)\right)\\
    \epsilon_{BE}&=\mathcal{O}\left( \frac{\delta\epsilon_{vN}}{\beta\log_2\left(\frac{6}{\delta}\right)}\right)
    \end{align}
    
    The first QFT step in \cref{alg:QSPQPE} can be implemented with $\text{Had}^{\otimes {r}}\ket{0^{{r}}}$ where $\text{Had}$ is the Hadamard gate \cite{Brassard_2002}. For the iQFT, we follow the standard cost assumption,  $\mathcal{O}({r}^2)=\mathcal{O}\left(\log_2^2\left(\frac{1}{\epsilon_{vN}}\right)\right)$ $1, 2$ qubit gates \cite{Nielsen_Chuang_2010}.

    The cost of block-encoding ${p}$ to precision $\frac{\epsilon_{BE}}{2\beta}$ is $\log({r})=\mathcal{O}\left(\log\log_2\left(\frac{1}{\epsilon_{vN}}\right)\right)$ extra qubits and so the overall $1, 2$-qubit cost is
    \begin{align}
{\mathcal{O}}&\left({r}^2+d {r}\log\left(\frac{2\beta}{1}\frac{\beta\log_2\left(\frac{6}{\delta}\right)}{\delta\epsilon_{vN}}\right)\right)\nonumber\\
&={\mathcal{O}}\left(\log^2_2\left(\frac{1}{\epsilon_{vN}}\right)+\log_2\left(\frac{1}{\epsilon_{vN}}\right)\right.\nonumber\\
&\left.\frac{\beta}{\epsilon_{vN}}\log_2\left(\frac{6}{\delta}\right)\cdot\log\left(\frac{2\beta^2\log_2\left(\frac{6}{\delta}\right)}{\delta\epsilon_{vN}}\right)\right)
    \end{align}
\end{proof}

\begin{corollary}[query-focused QPE]\label{cor:qpequery}
    Given access to $e^{iH}$, and otherwise make the same assumptions as \cref{thrm:qeequeryoptimal}. Then, one can produce a $\epsilon_{vN}$-close approximation of $\lambda_k$ with success probability $1-\delta$, using $\log(1/\epsilon_{vN})+n+5+\log\log(\epsilon_{vN}^{-1})$ qubits and 

    $$\tilde{\mathcal{O}}\left(\frac{\alpha}{\epsilon_{vN}}\log_2\left(\frac{1}{\delta}\right)\right)$$
    calls to controlled $e^{iH}$ and its inverse. The protocol also requires 
   \begin{align}
   {\mathcal{O}}&\left(\log_2^2\left(\frac{1}{\epsilon_{vN}}\right)+\frac{\alpha}{\epsilon_{vN}}\log_2\left(\frac{1}{\delta}\right)\right.\nonumber\\
    &\cdot \log_2\left(\frac{1}{\epsilon_{vN}}\right)\log\left(\frac{\alpha^2\log_2\left(\frac{1}{\delta}\right)}{\delta\epsilon_{vN}}\right)\nonumber
\end{align}
    $1, 2$-qubit gates
\end{corollary}
\begin{proof}
    We only have to account for the multiplicative overhead of going from $U\rightarrow BE(H)$, where the block-encoding is $\frac{\epsilon_{BE}}{2\cdot 2^{{r}}}=\frac{\epsilon_{BE}\epsilon_{vN}}{2}$-precise. 

      From \cref{lemma:gilyenBEoracle}, we can prepare a $\left(\frac{2\alpha}{\pi}, 2, \frac{\pi\epsilon_{BE}}{2(\alpha+2^{{r}})}\right)$ block-encoding of $H$. We have to use an additional qubit for the QSP+ circuit that implements this. With $\opnorm{{p}}=2^{{r}}\sum_{j=0}^{{r}-1}2^{j-{r}}\leq 2^{{r}}$, one can also prepare a $\left(\frac{2\cdot 2^{{r}}}{\pi}, \log\log(\epsilon_{vN}^{-1}), \frac{\pi\epsilon_{BE}}{2(\alpha+2^{{r}})}\right)$ block-encoding of ${p}$ with \cref{thrm:zhang7}. Using \cref{lemma:betensorlemma}, we have a $\left(\frac{4\alpha 2^{{r}}}{\pi^2}, 2+\log\log(\epsilon_{vN}^{-1}), \epsilon_{BE}\right)$ block-encoding of $B\otimes {p}$. The operation requires  $\mathcal{O}\left(\log\left(\frac{2^{{r}}+\alpha}{\epsilon_{BE}}\right)\right)$ calls to controlled $e^{iH}$. 

    $W(H\otimes {p})$,  a $\left(\frac{4\alpha {{r}}}{\pi^2}, 3+\log\log(\epsilon_{vN}^{-1}), \epsilon_{BE}\right)$ qubitized block-encoding, is built from $BE(H\otimes {p})$ with two uses of $BE(H\otimes {p})$ or $BE(H\otimes {p})^{\dag}$. Note that both the Hamiltonian simulation and block-encoding of $H$ require a single qubit for QSP rotations.

    The QFT costs are $\mathcal{O}\left(\log^2_2\left(\frac{1}{\epsilon_{vN}}\right)\right)$ as before. The block-encoding costs for $H$ are
    \begin{align}
 \cost{BE(H)}&={\mathcal{O}}\left(\log\left(\frac{\alpha^2}{\delta\epsilon_{vN}}\log_2\left(\frac{1}{\epsilon_{vN}}\right)\log_2\left(\frac{1}{\delta}\right)\right.\right.\nonumber\\
 &\left.\left.+\frac{\alpha}{\delta\epsilon_{vN}^2}\log_2\left(\frac{1}{\epsilon_{vN}}\right)\log_2\left(\frac{1}{\delta}\right)\right)\right)
    \end{align}
    Meaning that the overall number of queries to the block-encoding is 
    \begin{align}
       \mathcal{O}&\left(\frac{\alpha}{\epsilon_{vN}}\log_2\left(\frac{1}{\epsilon_{vN}}\right)\log_2\left(\frac{1}{\delta}\right)\right.\nonumber\\
        &\left.\cdot \log\left(\frac{\alpha}{\epsilon_{vN}}\log\left(\frac{1}{\epsilon_{vN\delta}}\right)\left[\frac{\alpha}{\delta}+\frac{1}{\epsilon_{vN}}\right]\right)\right)
    \end{align}
    Where in the lemma statement, $\tilde{\mathcal{O}}$ suppresses logarithmic factors of $\alpha, \log(1/ {\epsilon_{vN}})$, and $\log(1/\delta)$. The $1, 2$ qubit gate counting is derived similarly.
\end{proof}

\subsection{vN-QEE with  {LCP} oracles}
Now, we focus on QEE and explicitly assume that the oracle is prepared using an  {LCP} approach. This allows us to write the algorithm costs accounting for non-generic features of the Hamiltonian: the number of Pauli terms, the spectral gap, and the Hamiltonian norm.

\begin{lemma}[vN-QEE for LCP]\label{thrm:QEE-LPC}
Consider a Hamiltonian prepared as a LCP with $\left|P\right|$ terms and $\sum_j\alpha=\alpha$, assume we can prepare an approximation of the eigenstate corresponding to the $k$th eigenvalue $\lambda_k$ with gap $\Delta_k$, ie $\ket{\psi_0}=\sum c_j\ket{\lambda_j}$ where for $k\in [2^n]$, $c_{j}^2\leq \frac{\delta}{3\cdot 2^{n-1}}$ for all $j\neq k$. 

Then, one can prepare a $\epsilon_{vN}\in[0, 1]$ close approximation of the eigenvalue $\lambda_k$ with success probability $1-\mathcal{O}(\delta)$, $\delta\in(0, 1]$, $
\log(\epsilon_{vN}^{-1})+n+n_{anc}+2$ qubits,  {where}
 \begin{equation*}
     \Omega\left(\log_2\left|(1+\log(\epsilon_{vN}^{-1}))\log(\epsilon_{vN}^{-1})\right|\right)\leq n_{anc}.
 \end{equation*}

The $\mathcal{G}_{CpT}$ depth will be 
 \begin{align}
     \tilde{\mathcal{O}}&\left(\frac{\alpha}{\epsilon_{vN}}\log_2^2\left(\frac{1}{\delta}\right)\left|P\right|\left(1+\log\left(\frac{1}{\epsilon_{vN}}\right)\right)\right.\nonumber\\
     &\cdot \left.
     \left(n+\log\left(\frac{1}{\epsilon_{vN}}\right)\right)\log\left(\frac{\alpha^2\log_2\left(\frac{1}{\delta}\right)}{\delta\epsilon_{vN}}\right)\right.\nonumber\\
     &\left.\cdot \frac{\log(n_{anc})}{n_{anc}}+\frac{\alpha}{\epsilon_{vN}}\log_2^2\left(\frac{1}{\delta}\right)+\log_2^2\left(\frac{1}{\epsilon_{vN}}\right)\right)\nonumber.
 \end{align}
and the $\mathcal{G}_{CpT}$  gate complexity will be
 \begin{align}
     \mathcal{O}&\left(\frac{\alpha}{\epsilon_{vN}}\log_2^2\left(\frac{1}{\delta\epsilon_{vN}}\right)\left|P\right|\left(1+\log\left(\frac{1}{\epsilon_{vN}}\right)\right)\right.\nonumber\\
     &\left.\cdot \left[n+\log\left(\frac{1}{\epsilon_{vN}}\right)+\log\left(\frac{\alpha^2\log_2\left(\frac{1}{\delta}\right)}{\delta\epsilon_{vN}}\right)\right]\right)\nonumber
 \end{align}
\end{lemma}
\begin{proof}
    We account for the costs of building a block-encoding from the LCP. The encoded Hamiltonian will be the discretized pointer-system coupling with ${r}$ pointer qubits and norm as in \cref{eq:largenormmomentum},
\begin{align}
H&\otimes {p}=\sum \alpha_kP_k\otimes \sum_{j=1}^{{r}}2^{-j}\frac{1-\sigma_z^{(j)}}{2}\\
&=\sum_{j=1}^{{r}}\sum_{k}\frac{2^{-j}}{2}\alpha_kP_k+ \sum_{k, j=1}^{{r}}\frac{2^{-j}}{2}\alpha_kP_k\otimes \sigma_z^{(j)}\\
&=\sum_{k}{\alpha_{k}}'P_k\otimes I +\sum_{k, j}\alpha_{j, k}"P_k\otimes \sigma_z^{(j)}
\end{align}
which is another LCP, with $(1+{r})\left|P\right|$ terms. The sum of coefficients is now
\begin{align}
\sum& \frac{1}{2^{j+1}}\cdot \alpha+\sum_{k, j=1}^{{r}}\frac{\alpha_k}{2^{j+1}}\nonumber\\
&=\frac{\alpha}{2}\left(1-\left(\frac{1}{2}\right)^{{r}}\right)+\sum_k\frac{\alpha_k}{2}\left(1-\left(\frac{1}{2}\right)^{{r}}\right)\\
&\leq \alpha
\end{align}
In the large ${r}$-limit this will approach $\alpha$. Considering this worst case, we use \cref{thrm:zhang7} to prepare a $(\alpha, n_{anc},\epsilon_{BE}/\alpha)$ block-encoding of $H\otimes {p}$. The qubitized block-encoding is an $(\alpha, n_{anc}+1,\epsilon_{BE})$ block-encoding and requires  constant overhead calls to the first block-encoding.

Because we have decomposed the block-encoding costs into $\mathcal{G}_{CpT}$, we will do the same for the QET circuit, meaning that now the signal processing gates are prepared in $\mathcal{G}_{CpT}$ up to finite accuracy at least $\epsilon_{R}$. The accuracy of the QSP+ Hamiltonian simulation is now at least $\epsilon_{HS}=\epsilon_{QET}+2d\epsilon_{BE}+(2d+1)\epsilon_{R}$. We set $\epsilon_{R}\leq \epsilon_{QET}/(2d+1)$ and $\epsilon_{BE}\leq \epsilon_{QET}/2d$.
The success probability proceeds exactly as in \cref{thrm:qeequeryoptimal}, so we set $\epsilon_{HS}=\delta/3$ to get success probability $\geq 1-\delta$.  Now the bounds on $\epsilon_R, \epsilon_{BE}, d$ with respect to $\delta$ follow immediately. The block-encoding cost in $\mathcal{G}_{CpT}$ then follows from \cref{thrm:zhang7}, with 
  gate complexity
 \begin{align}
     \mathcal{O}&\left(d\left|P\right|\left(1+\log\left(\frac{1}{\epsilon_{vN}}\right)\right)\left[n\right.\right.\nonumber\\
     &\left.\left.+\log\left(\frac{1}{\epsilon_{vN}}\right)+\log\left(\frac{\alpha^2\log_2\left(\frac{1}{\delta}\right)}{\delta\epsilon_{vN}}\right)\right]\right)
 \end{align}
 and depth
  \begin{align}
     \tilde{\mathcal{O}}&\left(d\left|P\right|\left(1+\log\left(\frac{1}{\epsilon_{vN}}\right)\right)\left(n+\log\left(\frac{1}{\epsilon_{vN}}\right)\right)\right.\nonumber\\
     &\cdot \left.
     \log\left(\frac{\alpha^2\log_2\left(\frac{1}{\delta}\right)}{\delta\epsilon_{vN}}\right)\cdot \frac{\log(n_{anc})}{n_{anc}}\right).
 \end{align}
 In \cref{sec:qspnoisy} we work out the costs of a QET circuit in $\mathcal{G}_{CpT}$. Together with the query complexity and number of $1, 2$ qubit gates from \cref{thrm:qeequeryoptimal}, this directly gives the lemma statement. Now we can use \cite{matthewmeet2013} to ensure that the scaling of controlling the block-encodings is linear.
\end{proof}
\section{Discussion}
We have presented an efficient QEE algorithm, which can easily be modified to perform QPE, at the cost of a logarithmic overhead compared to the QPE lower bound \cite{mandetight2023}. We have focused on the standard QPE/QEE problem, meaning that we assume access to $\ket{\lambda}$, or a state exponentially close to it.

\Cref{thrm:QEE-LPC} then underscores that this efficiency holds when one adopts a common access model. The result illustrates how vN-QPE (and the underlying QET Hamiltonian simulation) complexity scales with physically relevant parameters $\alpha, \left|P\right|$ in practice. This is because while QSP+ templates provide an easy conceptual framework for algorithm development, important Hamiltonian parameters, usually $|P|, \Delta_k, \alpha$, are subsumed into block-encoding costs. This point is not truly novel, it can be derived straightforwardly from the complexity statements of block-encoding costs in for example \cite{Zhang_2024, gilyen_quantum_2019}. However, it is a key feature of algorithms on block-encodings data which deserves {prominence}.

\Cref{alg:QSPQPE} is amenable to simplifications that we have not discussed directly. First, for simple or particularly structured problems, then the block-encoding step might be done more cheaply than we have assumed (generic $e^{iH}$ oracles or some LCP). Regardless of whether the block-encodings are prepared with the methods we used, our work demonstrates that preparing and coupling the pointer system introduces minimal additional costs or assumptions to the problem.

Next, sometimes the initial QFT step in QPE can be replaced with a window function \cite{babbushencoding2018}. In analogy to classical signal processing, the window functions are designed to dampen spectral leaking, or the spreading of the probability distribution across all eigenstates of $U$ due to the QFT step or time-evolution. Window function-based QPE algorithms will have the same asymptotic query complexity as standard QPE, but could have a higher success probability, at the cost of a more expensive initial QFT step. Window function-QPE has been explicitly compared to QSVT-QPE, and numerically outperforms QSVT-based algorithms by a low constant factor \cite{greenaway2024casestudyqsvtassessment} in the test cases. The approach can be combined with ours, leading to a possible numeric boost to the success probability in $\gamma=1-\text{poly}(n)$ cases.

Finally, note that \cite{Magano_2022} shows that a variant of QPE, which explicates the depth-precision tradeoff inherent to all QPE algorithms, $\alpha$-QPE,  can also be formulated in QSVT. It is possible that by changing the subnormalization on $\Vec{p}$ {, one} could derive a similar result to $\alpha$-QPE.

\section*{Acknowledgments}
  Discussions with Tobias Osborne, Andreea Lefterovici, Ren\'e Schwonnek, and Martin Steinbach are gratefully acknowledged. We thank the anonymous reviewers for constructive feedback which improved the manuscript.

 We acknowledge the support of the Natural Sciences and Engineering Research Council of Canada (NSERC), PGS D - 587455 - 2024.

 Cette recherche a été financée par le Conseil de recherches en sciences naturelles et en génie du Canada (CRSNG), PGS D - 587455 - 2024. 

  This work was also directly or indirectly supported by the following project funding: Quantum Valley Lower Saxony; the Deutsche Forschungsgemeinschaft  project SFB
 1227 (DQ-mat);  Germany’s Excellence Strategy EXC-2123 QuantumFrontiers 390837967; the BMWK project ProvideQ, and by the BMBF projects QuBRA and ATIQ.

\bibliographystyle{plain}
\bibliography{qpebib}

\onecolumn\newpage
\appendix

\section{Block-encodings and qubitized block-encodings}\label{sec:qubitizedblockencodings}
 In QSP+ literature, the first proposals used the nomenclature "standard-form-encodings" which perfectly encoded some $H$ and used $\ket{G}$ to project to the subspace of $BE(H)$ containing it \cite{low_hamiltonian_2019}. From this definition, one defines a "qubitized standard-form-encoding operator" and then constructs a QET circuit.

\begin{definition}[qubitized block-encoding]
A  \textit{qubitized} block-encoding, $W(H)$, is a block-encoding as in \cref{def:blockencoding}, such that
 \begin{align}
     W(H)=\left(2\ket{G}\bra{G}-I\right)\otimes I \cdot S\cdot BE(H),
 \end{align}
 with the additional properties that
 $$\bra{G}S\cdot BE(H)S\cdot BE(H)\ket{G}=I_{rank(H)},$$
and 
$$\bra{G}S\cdot BE(H)\ket{G}=H.$$
Here,  $BE(H)$ is a block-encoding of $H$ in the same $\ket{G}$-basis and $S$
is a rotation that is implicitly defined in \cref{sec:qetoracles}.
\end{definition}

In \cite{low_hamiltonian_2019}, it was shown that a qubitized block-encoding can be built from $BE(H)$ using two queries to the block-encoding or its inverse and one additional qubit, when $\epsilon=0$. It is simple to show that a $\epsilon\neq 0$ block-encoding can be used to create an $\epsilon$-close qubitized block-encoding\textemdash for completeness this is done in \cref{sec:qetoracles}. We usually define $S$ so that the qubitized rotation basis is $\ket{+, G}$, and everywhere below we will use $\ket{G}=\ket{0^{\otimes n_{anc}}}$.

\begin{figure}[h]
\begin{tikzpicture}[node distance=1.5cm and 2cm, >=Stealth]

  % Define the styles for the nodes
  \tikzstyle{box} = [rectangle, draw, minimum width=3cm, minimum height=1cm, text centered]
  \tikzstyle{arrow} = [thick, ->]
  % Nodes with colors
  \node (input) [box, fill=c4!30] {Input $H$};
  \node (block) [box, fill=c5!30, below of=input] {Block-encoding $BE(W)$};
  \node (qubit) [box, fill=c3!30, below of=block] {Qubitized block-encoding $W(H)$};
  \node (signal) [box, fill=c4!30, below of=qubit] {Signal operator $CW$};

  % Curved arrows with text
  \draw [arrow] (input.east) to[out=0, in=0, looseness=1.2] node[midway, right] {$+n_{anc}$ qubits} (block.east);
  \draw [arrow] (block.east) to[out=0, in=0, looseness=1.2] node[midway, right] {$+1$ qubit} (qubit.east);
  \draw [arrow] (qubit.east) to[out=0, in=0, looseness=1.2] node[midway, right] {$+1$ qubit} (signal.east);
\end{tikzpicture}
\caption{Creating a QET signal operator from a Hamiltonian $H$. Note that most versions of QET modify the signal processing gates, instead of using $CW(H)$ \cite{lin_lecture_2022, low_hamiltonian_2019}. In QSP, one instead builds $CW$ from $e^{i\arccos(H)}$, which can be understood as a $(1, 0, 0)$ block-encoding. In QSVT, by convention one does not qubitize the block-encoding, and instead modifies the signal processing operators. Thus, the QSVT signal operator is simply $BE(H)$. }
\label{fig:qetoraclesketch}
\end{figure}

\subsection{Block-encodings from LCP Hamiltonians}
The standard way to prepare a block-encoding of a LCP is to construct two matrices, $ U_{PREP}$ and $U_{SELECT}$ \cite{childshslcu2012, kotharithesis}, 
\begin{align}
  U_{PREP} &:\ket{0^{\otimes \log_2(|P|)}}\rightarrow \frac{1}{\sqrt{\alpha}}\sum_{l\in [\left|P\right|]}\sqrt{\alpha_l}\ket{l},\label{eq:PREP}\\
    U_{SELECT}&=\sum_l\ket{l}\bra{l}\otimes P_l\label{eq:SELECT}\\
\end{align}
One can verify by explicit construction that $U_{PREP}^{\dag}U_{SELECT}U_{PREP}$ prepares a $\beta={\alpha}$ block-encoding of $H$. 

The construction of \cite{Zhang_2024} gives algorithms for approximately creating  $U_{SELECT}, U_{PREP}$, allowing for a variable number of ancillary qubits to be used. As in \cite{Zhang_2024}, we assume that $H$ has dimension $N=2^n$ and that $\log_2\left|P\right|$ is an integer\footnote{trivial to generalize by adding more terms with amplitude $0$ until $\left|P\right|$ is a power of $2$}. 
\begin{theorem}[Theorem 7 from \cite{Zhang_2024}]\label{thrm:zhang7}
    With $n_{anc}$ ancillary qubits where $\Omega\left(\log_2\left|P\right|\right)\leq n_{anc}\leq \mathcal{O}\left(2^n\left|P\right|\right)$, the $\left(1, n_{anc},  \epsilon\right)$ block-encoding of $H$ defined with a LCP such that $\alpha=1$ can be constructed with  $\mathcal{O}\left(\left|P\right|(n+\log(1/\epsilon))\right)$ count and  $\tilde{\mathcal{O}}\left(\left|P\right|n\log(1/\epsilon)\frac{\log n_{anc}}{n_{anc}}\right)$ depth of Clifford and $T$ gates, where $\tilde{\mathcal{O}}$ suppresses the doubly logarithmic factors of $n_{anc}$
\end{theorem}
 \Cref{thrm:zhang7} has the following trivial generalization for $\alpha\neq 1$. For a given $H$, consider the renormalized Hamiltonian $H'=\frac{H}{\alpha}$. $\opnorm{H'}=1$ and the construction in  \cite{Zhang_2024} prepares $BE(H'),$ encoding some Hamiltonian $\Tilde{H}$, which is $\epsilon$-close to $H'$, and therefore also $\epsilon$-close to $\frac{H}{\alpha}$. Thus,  $BE(H')$  is also an $\alpha\epsilon$-close block-encoding of $H$.

\subsection{Block-encodings from QPE oracles}
Recall that in QPE, one assumes access to $U=e^{iH}$, which in our setting will need to be converted into a block-encoding of $H$. A protocol from \cite{gilyen_quantum_2019} shows how to create\footnote{The original lemma statement requires $H$ to have a matrix norm of at most $1/2$; here we subnormalize $H\rightarrow \frac{H}{2\opnorm{H}}$} a $BE(H)$ from $U$. 
\begin{lemma}[Corollary 71 from \cite{gilyen_quantum_2019}]\label{lemma:gilyenBEoracle}
    Suppose that $U=e^{iH}$, where $H$ is a Hamiltonian of norm  $\opnorm{H}$.  Let $\epsilon\in\left(0, \frac{1}{2\opnorm{H}}\right]$, then we can implement a $\left(\frac{2\opnorm{H}}{\pi}, 2, \epsilon\right)$  block-encoding of $H$ with $\mathcal{O}\left(\log\left(\frac{1}{\epsilon}\right)\right)$ uses of controlled-U and its inverse, using $\mathcal{O}\left(\log\left(\frac{1}{\epsilon}\right)\right)$  two-qubit gates and using a single ancilla qubit.
\end{lemma}
Heuristically speaking, \cref{lemma:gilyenBEoracle} implements {$U\rightarrow i\log(U)$}. The proof is simple, and relies on two steps: (i) Show that $-iCU^{\dag}\left(ZX\otimes I\right)CU$ is a block-encoding of $\sin(H)$, where $C$ denotes a controlled operation. (ii) Implement a polynomial expansion of $\arcsin$ in QSVT, resulting in $BE(H)$.  

We refer the reader to \cite{gilyen_quantum_2019} for details on the functional approximation of $\arcsin$ and the bounds.

\subsection{Tensor products of block-encodings}
Finally, we need a  small lemma to show that block-encoding structure is preserved by the tensor product, \cref{lemma:betensorlemma}\footnote{This is essentially the same as a result found in \cite{Camps_2020}, but in our case the use of swap gates to needed to keep $\ket{G_A, G_B}=\ket{0^{n_A+n+B}}$ can be neglected.}.
 \begin{restatable}[Tensor products of block-encodings]{lemma}{tplemma}\label{lemma:betensorlemma}
    Given $BE(H_A), BE(H_B)$, respectively a $\left(\beta_A, n_A, \epsilon_A\right)$ block-encoding of Hamiltonian $H_A$ and a $\left(\beta_B, n_B, \epsilon_B\right)$ block-encoding of $H_B$. Say that $\opnorm{H_A}\leq \beta_A$, $\opnorm{H_B}\leq \beta_B$. Then, $BE(H_A)\otimes BE(H_B)$ is a $\left(\beta_A\beta_B, n_A+n_B, \beta_B\epsilon_A+\beta_A\epsilon_{B}\right)$ block-encoding of $H_A\otimes H_B$
\end{restatable}
\begin{proof}
Consider 
\begin{align}
   & \opnorm{H_A\otimes H_B-\beta_A\beta_B\bra{G_AG_B}BE(H_A)\otimes BE(H_B)\ket{G_AG_B}}\\
   &\leq  \opnorm{H_A\otimes H_B-\beta_A\bra{G_A}BE(H_A)\ket{G_A}\otimes H_B} \nonumber\\
   &+ \opnorm{\beta_A\bra{G_A}BE(H_A)\ket{G_A}\otimes H_B-\beta_A\beta_B\bra{G_AG_B}BE(H_A)\otimes BE(H_B)\ket{G_AG_B}}\\
    &\leq  \opnorm{H_A-\beta_A\bra{G_A}BE(H_A)\ket{G_A}}\cdot \opnorm{H_B}\nonumber\\
    &+ \opnorm{\beta_A\bra{G_A}BE(H_A)\ket{G_A}}\cdot \opnorm{H_B-\beta_B\bra{G_B}BE(H_B)\ket{G_B}}\\
    &\leq \epsilon_A\beta_B+\beta_A\epsilon_{B}
\end{align}
\end{proof}

\section{Qubitization and QET}\label{sec:qetoracles}

\textit{Qubitization} is the process of taking a block-encoding and transforming it into an iterate $W$ such that $W$ encodes the complex-exponential of the eigenvalues of $H$ in a direct product form. This can be done for any $(1, n_{anc}, 0)$ block-encoding $BE(H)$ with a constant number of calls to $BE(H), BE(H)^{\dag}$ and controlled versions. Below is a review of the standard reference \cite{low_hamiltonian_2019}. We could instead have followed \cite{lin_lecture_2022}, which presents a derivation of QET more closely in alignment with QSVT (promoting the signal processing operator to a controlled rotation, instead of promoting the block-encoding to a qubitized block-encoding).

Consider iterates of the form $W=\left(2\ket{G}\bra{G}-I\right)\cdot S BE(H)$, and for now ignore the definition of rotation $S$. Denote $\ket{G_{\lambda}}=\ket{G}\otimes \ket{\lambda}$, then one can show

\begin{equation}
    \ket{G_{\lambda}^{\perp}}=\frac{-\lambda\ket{G_{\lambda}}+W\ket{G_{\lambda}}}{\sqrt{1-|\lambda|^2}}.
\end{equation}
One can then define operations which act as the Pauli matrices in each $\ket{G_{\lambda}}$ subspace. That is, there exist $X_{\lambda}, Y_{\lambda}, Z_{\lambda}$ such that
\begin{align}
    X_{\lambda}\ket{G_{\lambda}}&=\ket{G_{\lambda}^{\perp}}\\
    Y_{\lambda}\ket{G_{\lambda}}&=i\ket{G_{\lambda}^{\perp}}\\
    Z_{\lambda}\ket{G_{\lambda}}&=\ket{G_{\lambda}}\\
\end{align}
And then finally $W$ is written as a direct product of these subspaces,
\begin{equation}
    W=\bigoplus\begin{bmatrix}
        \lambda & \sqrt{1-|\lambda|^2}\\
        \sqrt{1-|\lambda|^2} & \lambda
    \end{bmatrix}=\bigoplus_{\lambda}e^{-i\arccos(\lambda)Y_{\lambda}}.
\end{equation}
One can easily verify that $\cup_{\lambda}\{\ket{G_{\lambda}}, \ket{G_{\lambda}^{\perp}}\}$ form an orthonormal basis, with the following action on $W$
\begin{align}
    W\ket{G_{\lambda}}&=\begin{bmatrix}
        \lambda & \sqrt{1-|\lambda|^2}\\
        \sqrt{1-|\lambda|^2} & \lambda
    \end{bmatrix}\begin{bmatrix}
        \ket{G_{\lambda}}\\
        0
    \end{bmatrix}=\begin{bmatrix}
        \lambda\ket{G_{\lambda}}\\
        \sqrt{1-|\lambda|^2}\ket{G_{\lambda}^{\perp}}
    \end{bmatrix}\\
    W\ket{G_{\lambda}^{\perp}}&=\begin{bmatrix}
        \lambda & \sqrt{1-|\lambda|^2}\\
        \sqrt{1-|\lambda|^2} & \lambda
    \end{bmatrix}\begin{bmatrix}
    0\\
        \ket{G_{\lambda}^{\perp}}
    \end{bmatrix}=\begin{bmatrix}
        -\sqrt{1-|\lambda|^2}\ket{G_{\lambda}}\\
        \lambda\ket{G_{\lambda}^{\perp}}
    \end{bmatrix} \label{eq:WGlambdadag}.
\end{align}
The eigenvectors of $W$ are derived as linear combinations of \cref{eq:WGlambdadag}; the eigenvalues are $e^{i\Phi\mp i\arccos(\lambda)}$. Additionally, one can introduce a global phase to the operator; $e^{i\Phi}W$ has the same eigenvectors and eigenvalues $e^{i\Phi\mp i\arccos(\lambda)}$. Finally, the direct product carries through the exponentiation, so the generator of $W$ is $\bigoplus \arccos(\lambda)Y_{\lambda}$.

With $W$ and its eigenspectrum, one can define a phased iterate $W_{\phi},$ which applies a sequence of phases on $W$ that ultimately represents a polynomial in the eigenvalues. This will be the QET circuit. First, define operation
\begin{equation}
    Z_{\phi}=\bigoplus \begin{bmatrix}
        e^{-i\phi} & 0\\
        0 & -1\\
    \end{bmatrix}=\bigoplus_{\lambda}e^{-\phi/2}e^{-i\phi/2Z_{\lambda}},
\end{equation}
and then consider the product operation which builds up polynomials of $H$.\footnote{We have suppressed a translation in each $\phi_j$ for convenience, see \cite{low_hamiltonian_2019} for the detailed construction.}
\begin{align}
    W_{QET}=\prod_{\{\phi_j\}}Z_{\phi_j}WZ_{-\phi_j}.
\end{align}
This isn't a real circuit yet, but it can be implemented with one additional qubit and a series of controlled $Z$ operations about $G$ \cite{low_hamiltonian_2019, lin_lecture_2022}. In QSP+ language, the signal processor has been promoted to $\left(2\ket{G}\bra{G}-I\right))\left(e^{i(\phi_j)}\otimes I\right)\left(2\ket{G}\bra{G}-I\right)$ , and $W$ is the signal operator.

We now promote the $W$ operation to a controlled operation on $C_{+}W$ for even polynomials, which is single-ancilla QSP in \cite{low_hamiltonian_2019}. In this case, a global phase is allowed on $W$, so $e^{i\Phi}W$ is the qubitized block-encoding used. The QET circuit for Chebyshev polynomials is then
\begin{align}
    U_{QET}=(e^{i\frac{\phi_0}{2}Z}\otimes I)\prod_{\{\phi_j\}}(e^{i\frac{\phi_j}{2}Z}\otimes I)C_{-}W(e^{-i\frac{\phi_j-\phi_{j+1}}{2}Z}\otimes I)C_{+}W^{\dag}(e^{-i\frac{\phi_{j+1}}{2}Z}\otimes I).
\end{align}
One can actually implement more general rotations than $Z_{\phi}$, or subsume the control basis change into $Z$, resulting in a circuit with $C_1W$  and gates with $1$-qubit parameterized rotations $R(\phi)$. The QSP convention from \cite{Haah2019product}  immediately extends to the direct product notation for QET, and the most general QSP convention of \cite{motlagh_generalized_2023} allows any $U(2)$ rotation to be used  (and is extendable to QET or QSVT). One must take care that the QSP+ circuit design of choice admits polynomials of the desired form.

\subsection{Approximate qubitized block-encodings}
Given $\left(1, n_{anc}, 0\right)$ block-encoding $BE(H)$, one could try to identify an $S$ such that
\begin{align}
    \bra{G}S\cdot BE(H)\ket{G}=H & \text{ and } \bra{G}S\cdot BE(H)\cdot S\cdot BE(H)\ket{G}=I_{rank(H)}.
\end{align}
More generally, \cite{low_hamiltonian_2019} showed that one can construct qubitized block-encoding $W(H)$ for any block-encoding $BE(H)$ using an additional ancilla qubit. Specifically, 
\begin{equation}
 W(H)=\left(\sigma_X\otimes I_{n_{anc}+rank(BE)}\right)C_1BE(H)^{\dag}C_0BE(H).\label{eq:qubitizedBE}
\end{equation}
$W(H)$ will project around the subspace defined by $\ket{G'}=\ket{+}\otimes\ket{G}$.

We now verify that the construction above extends to $\left(\beta, n_{anc}, \epsilon\right)$ block-encodings. In this case, there is another Hamiltonian, $H'$, which is $\epsilon$-close to $H/\beta$ and $BE(H)$ is a perfect block-encoding for $H'$. Then $H'$ admits a qubitized oracle, $W(H')$, such that 
\begin{align}
    \opnorm{H-\beta\bra{G'}W(H')\ket{G'}}= \opnorm{H-\beta H'}\leq \epsilon.
\end{align}
Thus, $W(H')$ is a $\left(\beta, n_{anc}+1, \epsilon\right)$ block-encoding for $H$.

\section{QET error}
We consider three sources of error in QET circuits. First, in QET one usually considers a polynomial which is  $\epsilon_{approx}\in[0,1)$ close to a desired function $f$ in the $l_{\infty}$-norm, and then has some controllable\footnote{Controlling circuit decomposition error is rigorously discussed in \cite{Haah2019product, ni2024fastphasefactorfinding} and some optimization-based pre-processing methods bypass the problem completely \cite{dong2023robustiterativemethodsymmetric}. For a review see \cite{skelton2025hitchhikersguideqsppreprocessing}.} error from the QSP-processing step to derive a QSP circuit from the input polynomial. We jointly capture the functional and decomposition error with $\epsilon_{QET}\in[0, 1)$, and one can say that the QET circuit implements a polynomial $P'$ which is $\epsilon_{QET}$-close to $f$. After post-selecting the ancilla qubits from the QET rotation qubit and the block-encoding ancillary qubits, one arrives at 
\begin{align}
&\opnorm{\sum_{\lambda}P'\left(\frac{\lambda}{\alpha}\right)\ket{\lambda}\bra{\lambda}-\sum_{\lambda}f\left(\frac{\lambda}{\alpha}\right)\ket{\lambda}\bra{\lambda}}\nonumber\\
&\leq \opnorm{\sum\epsilon_{QET} \ket{\lambda}\bra{\lambda}}\leq \epsilon_{QET}.
\end{align}
Thus, $\epsilon_{QET}$ is the QET circuit error if gates are implemented perfectly and a perfect qubitized block-encoding is used. 

Next, one must consider how  the error in the signal operator,  $\epsilon_{BE}\in[0, 1),$ propagates through the QET circuit. If $\epsilon_{BE}$ is the error in the $\mathcal{G}_{CpT}$ decomposition, then for consistency we also account for the costs of decomposing the signal processing gates in $\mathcal{G}_{CpT}$, implemented up to some precision $\epsilon_{R}\in[0, 1)$. 

In \cref{sec:gateerror}, we show that $\epsilon_R, \epsilon_{BE}$ can be jointly managed and that the distance between the approximately-implemented QET circuit $\tilde{U}_{QET}$ and the ideally-implemented circuit $U_{QET}$ is constrained by
\begin{equation}
    \opnorm{U_{QET}-\tilde{U}_{QET}}\leq  d\left(\epsilon_{R}+\epsilon_{BE}\right)+\epsilon_R.
\end{equation}

The error arising from the approximate implementation and the functional approximation are additive, so overall, 
\begin{align}
    &\opnorm{\sum f(\lambda)\ket{\lambda}\bra{\lambda}-\bra{G',a_{QET}}\tilde{U}_{QET}\ket{G',a_{QET}}}\nonumber\\
    \quad &\leq \epsilon_{QET}+2d\epsilon_{BE}+\left(2d+1\right)\epsilon_R.
\end{align}

\subsection{QET gate error}
\label{sec:gateerror}
The algorithms in the main text rely on being able to bound the propagation of gate error through QET circuits. This can easily be derived for any QSP+ circuit; for completeness we do so in \cref{lemma:approxqsperrorbound}. A QSP+ circuit encoding an degree-$d$ Laurent polynomial, where $d$ is even, has the following general form
\begin{equation}
    W_{QSP+}=R_0\prod_{d} CW\cdot\left(R_{2j-1}\otimes I\right)\cdot  CW^{\dag}\cdot \left( R_{2j}\otimes I\right)\label{eq:qsppluscircuit}.
\end{equation}
The odd-degree circuit has an analogous description with an additional $ CW\cdot\left(R_{2n+1}\otimes I\right)$ gate outside of the product. 

Say that signal operator(s) $CW$, $CW^{\dag}$, are implemented by unitaries $C\Tilde{W}$, $CW^{\dag}$ with some error at most $\epsilon_W$, and the signal processing operators $R_j$ are implemented by unitaries $\Tilde{R}_j$ with some error at most $\epsilon_{R}$. In this analysis, it is not important whether $\epsilon_R, \epsilon_W$ arise from a limited accuracy decomposition into some convenient elementary gate set, or because the oracle and/or QET circuit are derived to limited accuracy.

\begin{lemma}[Accuracy of an approximately implemented QSP+]\label{lemma:approxqsperrorbound}
Say that signal operators $CW, CW^{\dag}$ are implemented with accuracy $\epsilon_W\in[0, 1)$ and signal processing operators $\{R_j\}$ are implemented with accuracy $\epsilon_R\in[0, 1)$. Then, the QSP+ implementation remains $\mathcal{O}\left(n\epsilon_R+n\epsilon_W\right)$ close to the ideal circuit.
\end{lemma}

\begin{proof}
The quantity one must bound is the difference between the noisy implementation and the QSP+ circuit,
\begin{align}
    \opnorm{W_{QSP+}-\Tilde{W}_{QSP+}}&\leq \opnorm{\prod_{[d]}CW...-\prod_{[d]}C\Tilde{W}...}\\
    &+\opnorm{R_0-\Tilde{R}_0}\cdot \opnorm{\prod_{[d]}CW...}\\
    &\leq \opnorm{\prod_{[d]}CW...-\prod_{[d]}C\Tilde{W}...}+\epsilon_R.
\end{align}

The difference between one "cycle" $CWR_{2j-1}CW^{\dag}R_{2j}$ and its noisy implementation will be
\begin{align}
    &\opnorm{CWR_{2j-1}CW^{\dag}R_{2j}-C\Tilde{W}\Tilde{R}_{2j-1}C\Tilde{W}^{\dag}\Tilde{R}_{2j}}\\
    &\leq \epsilon_W+\left[\epsilon_R+\left(\epsilon_W+\epsilon_R\right)\right]\\
    &=2\epsilon_R+2\epsilon_{W}
\end{align}
After every iteration's error is accounted for,
\begin{align}
    \opnorm{W_{QSP+}-\Tilde{W}_{QSP+}}&\leq \left(g(\epsilon_W, \epsilon_R)+\opnorm{\prod_{j=2}^{d}CW...-\prod_{j=2}^{d}C\Tilde{W}...}\right)+\epsilon_R\\
    &\leq {d}\left(2\epsilon_R+2\epsilon_{W}\right)+\epsilon_R\\
    &=2d\epsilon_W+(2d+1)\epsilon_R
\end{align}
\end{proof}

\subsection{Approximate QET circuit depth}\label{sec:qspnoisy}
We now consider the $\mathcal{G}_{CpT}$ complexity and depth of a QET circuit. The Toffoli count of one-qubit gates can be found in  \cite{kliuchnikov2013fastefficientexactsynthesis, gosset2013algorithmtcount} for exact decompositions or in \cite{kliuchnikovasymptotically2013, kliuchnikovpractical2016, selingerefficient2015} for  $\epsilon_R$-close approximations.

We will assume access to a $\left(\beta, n_{anc}, \epsilon_{BE}\right)$  block-encoding of $H$ with some decomposition into Clifford and $T$ gates, $\mathcal{C}(BE(H))$. From $BE(H)$, $W(H)$ is given in \cref{eq:qubitizedBE}, which involves the cost of two controlled block-encodings and one Clifford gate. Then, the signal operator is the controlled version of $W(H)$, and we will assume that $W(H)$ and $W(H)^{\dag}$ have the same cost.

From \cite{matthewmeet2013}, the  $\mathcal{G}_{CpT}$  cost of $CV$ for any unitary $V$ will be linear in $\mathcal{C}(V)$, using at most one ancilla qubit. With this construction, $\cost{W(H)}$ has a $\mathcal{O}\left(\mathcal{C}(BE(H))\right)$ complexity and depth in $\mathcal{G}_{CpT}$. The signal processing oracle, $CW(H)$, is a controlled gate which can be prepared with the same process, so $\cost{CW}=\cost{BE(H)}$  in $\mathcal{G}_{CpT}$, for both the depth and the complexity. That is, the signal oracle $CW$ (or $CW^{\dag}$) will need at most two ancilla qubits in addition to $n+n_{anc}+2$ qubits for the $CW$ gate, but will remains linear in the costs of $(BE(H))$.

Because QET signal processing gates are in $SU(2)$ for our implementation \cite{Haah2019product}, they can be prepared to accuracy $\epsilon_R$ with  $\mathcal{O}\left(\log(2/\epsilon_R)\right)$ $\mathcal{G}_{CpT}$ gates and no ancillary qubits \cite{selingerefficient2015}. For a single-qubit rotation, this is also the $\mathcal{G}_{CpT}$ depth.

 Overall, one requires $\log_2(\text{rank}(H))+n_{anc}+2$ qubits for the QET circuit, and up to two ancillary qubits used in the $\mathcal{G}_{CpT}$ decomposition. From \cref{eq:qsppluscircuit}, the QET circuit uses $\mathcal{O}\left((2d+1)(\log(1/\epsilon_R)\right)$ $\mathcal{G}_{CpT}$ gates for the signal processing gates and $\mathcal{O}\left(d\mathcal{C}(BE(H))\right)$ gates for the signal oracle steps overall, so that
\begin{equation}
    \mathcal{C}\left(W_{QSP+}(H)\right)=\mathcal{O}\left(d\mathcal{C}(BE(H))+(2d+1)(\log(1/\epsilon_R)\right)\label{eq:qet-cpt-scaling}.
\end{equation}
These are all consecutive operations, and so discounting circuit optimization techniques, \cref{eq:qet-cpt-scaling} is also the $\mathcal{G}_{CpT}$ depth. 

\section{Example and Circuits}\label{sec:example}

We take $H=\omega \sigma_X$ as an example Hamiltonian, for some $\omega<1$. This system has two eigenstates $\omega, -\omega$, eigenvectors $\ket{\pm}=\frac{1}{\sqrt{2}}\left(\ket{0}+\ket{1}\right)$, spectral gap $\Delta_0=2\omega$, and spectral norm $\alpha=\omega$. We will use a pointer system with $2$ qubits, meaning that $\epsilon_{vN}\geq \frac{1}{4}$. 
The dicretized pointer system is
\begin{align}
    p=\frac{1-\sigma_z^{(0)}}{2}\otimes I+I\otimes 2\frac{I-\sigma_Z^{(1)}}{2}
\end{align}
and the discretized time-evolution is given by
\begin{align}
    e^{-it\omega \sigma_X\otimes p}.
\end{align}
The initial state will be 
\begin{align}
    \ket{\lambda_k, x(0)}&=\ket{+}\otimes\frac{1}{{2}}\left(\ket{00}+\ket{01}+\ket{10}+\ket{11}\right)\\
    &=\ket{+}\otimes \ket{+}\otimes \ket{+}.
\end{align}
Below, we specialize in the case where $\lambda_k=\omega$. 

Evolved for a time $t$, we have
\begin{align}
    e^{-it\omega \sigma_X\otimes p}\ket{+}\otimes \ket{+}\otimes \ket{+}&=\frac{1}{2}\ket{+}\left(\ket{00}+e^{-it\omega(0+\frac{1}{2})}\ket{01}+e^{-it\omega}\ket{10}+e^{-it\omega(1+\frac{1}{2})}\ket{11}\right)\\
    &=\frac{1}{2}\ket{+}\left(\ket{00}+e^{-it\omega\frac{1}{2}}\ket{01}+e^{-it\omega}\ket{10}+e^{-it\omega\frac{3}{2}}\ket{11}\right)
\end{align}
The iQFT results in
\begin{align}
    iQFTe^{-i\omega \sigma_X\otimes p}\ket{+}\otimes \ket{+}\otimes \ket{+}&=\frac{1}{2\cdot {2}}\ket{+}\left(\sum_{x=0}^{3}e^{i\frac{\pi x\cdot 0}{2}}\ket{x}+e^{-it\omega\frac{1}{2}}\sum_{x=0}^{3}e^{i\frac{\pi x\cdot 1}{2}}\ket{x}\right.\nonumber\\
    &\left.+e^{-it\omega}\sum_{x=0}^{3}e^{i\frac{\pi x\cdot 2}{2}}\ket{x}+e^{-it\omega\frac{3}{2}}\sum_{x=0}^{3}e^{i\frac{\pi x\cdot 3}{2}}\ket{x}\right)\\
     &=\frac{\ket{+}}{4}\otimes \left[\left(1+e^{-i\frac{t\omega}{2}}+e^{-i{t\omega}}+e^{-i\frac{3t\omega}{2}}\right)\ket{00}+\right.\nonumber\\
     &\left. ...+\left(1+e^{i\frac{3\pi}{2}}e^{-i\frac{t\omega}{2}}+e^{i3\pi}e^{-i{t\omega}}+e^{i\frac{9\pi}{2}}e^{-i\frac{3t\omega}{2}}\right)\ket{01}\right].
\end{align}
The success probability of some $x\in\{0, 1, 2, 3\}$ is
\begin{align}
\left|\kappa_x\right|&=\frac{1}{16}\frac{\sin^2\left[\pi \left(x-4{\omega}\right)\right]}{\sin^2\left[\frac{\pi}{4} \left(x-4{\omega}{}\right)\right]}.
\end{align}
In \cref{fig:succcprobwithr}, we show that the four success probabilities spike in the expected regions of $w\in[0, 1]$. The success probability is still relatively low in our example, but increasing $r$ will improve it.

\begin{figure}
    \centering
\definecolor{c4}{RGB}{164, 166, 46}
\definecolor{c3}{rgb}{0.13, 0.26, 0.12}
\definecolor{c2}{rgb}{0.8, 0.47, 0.13}
\definecolor{c1}{rgb}{0.0, 0.0, 1.0}
\begin{tikzpicture}
\begin{axis}[
    width =0.8*\linewidth,
height=0.45*\linewidth,
    xlabel={$\omega$},
    ylabel={$\left|\kappa_x\right|$},
    xmin=0, xmax=1,
    ymin=-0.05, ymax=1.05,
    xtick={0, 0.25, 0.5, 0.75, 1.0},
    xticklabels={$0$, $\tfrac{1}{4}$, $\tfrac{1}{2}$, $\tfrac{3}{4}$, $1$},
    ytick={0, 0.2, 0.4, 0.6, 0.8, 1.0},
    grid=both,
    grid style={line width=0.3pt, draw=gray!30},
    major grid style={line width=0.5pt, draw=gray!60},
    legend pos=outer north east,
    legend style={font=\small},
    tick label style={font=\small},
    label style={font=\small},
    title style={font=\small, yshift=4pt},
    samples=1000,
    smooth,
]
 
\addplot[
    color=c1,
    line width=1.4pt,
    domain=0.001:0.249,
] {(sin(deg(4*pi*x)))^2 / (16 * (sin(deg(pi*x)))^2)};
\addplot[
    color=c1,
    line width=1.4pt,
    domain=0.251:0.499,
    forget plot,
] {(sin(deg(4*pi*x)))^2 / (16 * (sin(deg(pi*x)))^2)};
\addplot[
    color=c1,
    line width=1.4pt,
    domain=0.501:0.749,
    forget plot,
] {(sin(deg(4*pi*x)))^2 / (16 * (sin(deg(pi*x)))^2)};
\addplot[
    color=c1,
    line width=1.4pt,
    domain=0.751:0.999,
    forget plot,
] {(sin(deg(4*pi*x)))^2 / (16 * (sin(deg(pi*x)))^2)};
\addplot[color=c1, only marks, mark=square, mark size=2pt, forget plot] coordinates {(0,1)};
\addlegendentry{$x=0$}
 
\addplot[
    color=c2,densely dashed,
    line width=1.4pt,
    domain=0.001:0.249,
] {(sin(deg(pi*(1-4*x))))^2 / (16 * (sin(deg(pi*(1-4*x)/4)))^2)};
\addplot[
    color=c2,densely dashed,
    line width=1.4pt,
    domain=0.251:0.499,
    forget plot,
] {(sin(deg(pi*(1-4*x))))^2 / (16 * (sin(deg(pi*(1-4*x)/4)))^2)};
\addplot[
    color=c2,densely dashed,
    line width=1.4pt,
    domain=0.501:0.749,
    forget plot,
] {(sin(deg(pi*(1-4*x))))^2 / (16 * (sin(deg(pi*(1-4*x)/4)))^2)};
\addplot[
    color=c2,densely dashed,
    line width=1.4pt,
    domain=0.751:0.999,
    forget plot,
] {(sin(deg(pi*(1-4*x))))^2 / (16 * (sin(deg(pi*(1-4*x)/4)))^2)};
\addplot[color=c2, only marks, mark=x, mark size=2pt, forget plot] coordinates {(0.25,1)};
\addlegendentry{$x=1$}
 
\addplot[
    color=c3, dotted,
    line width=1.4pt,
    domain=0.001:0.249,
] {(sin(deg(pi*(2-4*x))))^2 / (16 * (sin(deg(pi*(2-4*x)/4)))^2)};
\addplot[
    color=c3,dotted,
    line width=1.4pt,
    domain=0.251:0.499,
    forget plot,
] {(sin(deg(pi*(2-4*x))))^2 / (16 * (sin(deg(pi*(2-4*x)/4)))^2)};
\addplot[
    color=c3,dotted,
    line width=1.4pt,
    domain=0.501:0.749,
    forget plot,
] {(sin(deg(pi*(2-4*x))))^2 / (16 * (sin(deg(pi*(2-4*x)/4)))^2)};
\addplot[
    color=c3,dotted,
    line width=1.4pt,
    domain=0.751:0.999,
    forget plot,
] {(sin(deg(pi*(2-4*x))))^2 / (16 * (sin(deg(pi*(2-4*x)/4)))^2)};
\addplot[color=c3, only marks, mark=o, mark size=2pt, forget plot] coordinates {(0.5,1)};
\addlegendentry{$x=2$}
 
\addplot[
    color=c4,dashed,
    line width=1.4pt,
    domain=0.001:0.249,
] {(sin(deg(pi*(3-4*x))))^2 / (16 * (sin(deg(pi*(3-4*x)/4)))^2)};
\addplot[
    color=c4,dashed,
    line width=1.4pt,
    domain=0.251:0.499,
    forget plot,
] {(sin(deg(pi*(3-4*x))))^2 / (16 * (sin(deg(pi*(3-4*x)/4)))^2)};
\addplot[
    color=c4,dashed,
    line width=1.4pt,
    domain=0.501:0.749,
    forget plot,
] {(sin(deg(pi*(3-4*x))))^2 / (16 * (sin(deg(pi*(3-4*x)/4)))^2)};
\addplot[
    color=c4,dashed,
    line width=1.4pt,
    domain=0.751:0.999,
    forget plot,
] {(sin(deg(pi*(3-4*x))))^2 / (16 * (sin(deg(pi*(3-4*x)/4)))^2)};
\addplot[color=c4, dashed, only marks, mark=*, mark size=2pt, forget plot] coordinates {(0.75,1)};
\addlegendentry{$x=3$}
 
\end{axis}
\end{tikzpicture}
    \caption{Success probability of the $r=2$ vN-QEE for $t= 2\pi$}
    \label{fig:succcprobwithr},
\end{figure}
For example, say we measure in the computational basis and get $\ket{01}$, then the estimate will be
\begin{align}
    \left|\tilde{\omega}_k\right|= \frac{1}{2}\pm \frac{1}{4}.
\end{align}
 
To implement this with vN-QEE, we first would need to block-encoding $H\otimes \Vec{p}$. The LCP block-encoding is prepared through access to $U_{SELECT}, U_{PREP}$ as defined in \cref{eq:SELECT}, \cref{eq:PREP}, leading to $BE(H)$ in \cref{fig:BE}. We then qubitize the block-encoding, leading to $W(H)$. vN-QEE is performed with \cref{fig:vnqee}, with the Hamiltonian simulation subroutine given in \cref{fig:qspbasiccircuit}.

For this Hamiltonian, vN-QEE has no advantage over textbook QPE using to estimate $\omega$.  The textbook QPE has controlled powers of  $U^{j}$ sandwiched between a QFT and iQFT step; the discretized vN-QPE essentially replaces the controlled powers with a Hamiltonian simulation subroutine as in \cref{fig:qspbasiccircuit} and \cref{fig:vnqee}. Using QET for the Hamiltonian simulation, one must  build controlled block-encodings and one-qubit rotations. 

However, for a commuting Hamiltonian like $\omega \sigma_X$, $e^{i\omega}$ is as plausible to prepare as $e^{i j\omega \sigma_X}$, meaning that preparing the block-encoded system-pointer system instead of $e^{i j\omega \sigma_X}$ will not give a computational advantage. Where vN-QEE could provide an advantage is when it is not clear how to implement $U^{j}$ without overhead costs, such as for non-commuting Hamiltonians with many terms.

\begin{figure}
\centering
\resizebox{0.8\textwidth}{!}{
\begin{tikzpicture}
\begin{yquant}
qubit {$\ket{0}$} be[1];
qubit {$\ket{0}$} r[2];
qubit {$\ket{0}$} syst[1];
box {$U_{PREP}$} be;
box {$\sigma_X$} syst;
box {$I-\sigma_Z$} (r[0]) | ~be;
box {$I-\sigma_Z$} (r[1]) | be;
% box {$\sigma_X\otimes I$} (syst, r)| ~be;
% box {$\sigma_X\otimes \sigma_Z$} (syst, r)| be;
box {$U_{PREP}^{\dag}$} be;
\end{yquant}
\end{tikzpicture}=\begin{tikzpicture}
\begin{yquant}
qubit {$\ket{0}$} be[1];
qubit {$\ket{0}$} r[2];
qubit {$\ket{0}$} syst[1];
box {$BE(H\otimes \Vec{p})$} (be, syst, r);
\end{yquant}
\end{tikzpicture}
}
\caption{Basic LCP block-encoding for $H\otimes \Vec{p}=\frac{\omega}{4}\sigma_X\otimes p$, $r=2$}
    \label{fig:BE}
\end{figure}

\begin{figure}
\centering
\resizebox{0.8\textwidth}{!}{
\begin{tikzpicture}
\begin{yquant}
qubit {$\ket{+}$} q[1];
qubit {$\ket{0}$} be[1];
qubit {$\ket{0}$} r[2];
qubit {$\ket{0}$} syst[1];

box {$BE(H)$} (be, syst, r) | ~q;
box {$BE(H)^{\dag}$} (be, syst, r) | q;
box {$\sigma_X$} q;
\end{yquant}
\end{tikzpicture}=\begin{tikzpicture}
\begin{yquant}
qubit {$\ket{+}$} q[1];
qubit {$\ket{0}$} be[1];
qubit {$\ket{0}$} r[2];
qubit {$\ket{0}$} syst[1];
box {$W(H\otimes \Vec{p})$} (q, be, syst, r);
\end{yquant}
\end{tikzpicture}
}
\caption{Qubitized block-encoding of $H\otimes \Vec{p}$}
    \label{fig:qubitBE}
\end{figure}

\begin{figure}
\centering
\yquantdefinebox{dots}[inner sep=0pt]{$\dots$}
\resizebox{\textwidth}{!}{
\begin{tikzpicture}
\begin{yquant}
qubit {$\ket{+}$} j[1];
nobit ellipsis;
qubit {$\ket{+}$} q[1];
qubit {$\ket{0}$} be[1];
qubit {$\ket{0}$} r[2];
qubit {$\ket{0}$} b[1];

box {$e^{i\phi_{2d}}$} j;
box {$W^{\dag}(H)$}  (b, be, q, r)| ~j;
box {$e^{i\phi_{2d-1}}$} j;
box {$W(H)$} (b, be, q, r)| j;
box {$e^{i\phi_{2d-2}}$} j;
box {$W^{\dag}(H)$}  (b, be, q, r)| ~j;
box {$e^{i\phi_{2d-3}}$} j;
box {$W(H)$}  (b, be, q, r)| j;

barrier (-);
[style={draw=none}]
box {$\quad\cdots\quad$} j;
barrier (-);

box {$e^{i\phi_{2}}$} j;
box {$W^{\dag}(H)$}  (b, be, q, r)| ~j;
box {$e^{i\phi_{1}}$} j;
box {$W(H)$} (b, be, q, r)| j;
box {$e^{i\phi_0}$} j;
box {$\text{Had}$} j;
\end{yquant}
\end{tikzpicture}
}
\caption{Basic QSP circuit for operator $W(H)$, preparing degree $d$ Laurent
polynomial $\mathcal{P}_{HS} : \lambda \rightarrow |z| < 1$ given set of
unitaries $\{e^{i\phi_j}\} \in SU(2)^{2d+1}$.}
\label{fig:qspbasiccircuit}
\end{figure}

\begin{figure}
\centering
\resizebox{0.8\textwidth}{!}{
\begin{tikzpicture}
\begin{yquant}
qubit {$\ket{+}$} j[1];
qubit {$\ket{+}$} q[1];
qubit {$\ket{be}$} be[1];
qubits {$\ket{r}$} r[1];
qubits {$\ket{n}$} b[1];
box {$\text{Had}$} j;
box {$\text{Had}$} r;
box {$U_{QET}\left(H\otimes \Vec{p}\right)$} (j, q, be, b, r);
box {$iQFT$} r;
measure j;
measure r;
\end{yquant}
\end{tikzpicture}
}
\caption{the vN-QEE or vN-QPE circuit, implementing \cref{alg:QSPQPE}. The registers are respectively $n=\lceil\log(\text{rank}(H))\rceil$ for the "system" qubits, $r$ for the pointer system, $be$ for the qubits introduced for the block-encoding qubits, and $\ket{+}$ for both the qubitization qubit and the QSP rotation qubit.}
    \label{fig:vnqee}
\end{figure}
\end{document}